\documentclass[apj]{emulateapj}

\usepackage[usenames,dvipsnames,svgnames,hyperref,amsmath]{xcolor}
\usepackage{soul,threeparttable}
\usepackage{color}
\usepackage{natbib}

\defcitealias{lee13}{LY13}
\newcommand{\subfind}{\textsc{Subfind}}
\newcommand{\mpgrafic}{\textsc{MPgrafic}}
\newcommand{\grafic}{\textsc{Grafic-1}}
\newcommand{\ysamtm}{\textsc{ySAMtm}}
\newcommand{\ysam}{\textsc{ySAM}}
\newcommand{\gadgett}{\textsc{Gadget-2}}

\shorttitle{Formation and assembly history of galaxies}
\shortauthors{Jaehyun Lee and Sukyoung K. Yi}

\begin{document}
\title{Formation and assembly history of stellar components in galaxies as a function of stellar and halo mass}
\author{Jaehyun Lee\altaffilmark{1} and Sukyoung K. Yi\altaffilmark{2}}
\altaffiltext{1}{Korea Astronomy and Space Science Institute, 776, Daedeokdae-ro, Yuseong-gu, Daejeon 34055, Republic of Korea; syncphy@gmail.com}
\altaffiltext{2}{Department of Astronomy and Yonsei University Observatory, Yonsei University, Seoul 03722, Republic of Korea}

\begin{abstract}

Galaxy mass assembly is an end product of structure formation in the $\Lambda$CDM cosmology. As an extension of Lee \& Yi (2013), we investigate the assembly history of stellar components in galaxies as a function of halo environments and stellar mass using semi-analytic approaches. In our fiducial model, halo mass intrinsically determines the formation and assembly of the stellar mass. Overall, the ex situ fraction slowly increases in central galaxies with increasing halo mass but sharply increases for $\log M_{*}/M_{\odot}\gtrsim11$. A similar trend is also found in satellite galaxies, which implies that mergers are essential to build stellar masses above $\log M_{*}/M_{\odot}\sim11$. We also examine the time evolution of the contribution of mass growth channels. Mergers become the primary channel in the mass growth of central galaxies when their host halo mass begins to exceed $\log M_{200}/M_{\odot}\sim13$. However, satellite galaxies seldom reach the merger-dominant phase despite their reduced star formation activities due to environmental effects. 

\end{abstract}

\section{Introduction}

Modern theories for galaxy formation and evolution suggest that galaxies are formed in highly over-dense regions, namely haloes, while cosmological structures are built up via smooth matter accretion or the coalescence of haloes. Thus, some galaxies are eventually involved in hierarchical mergers in the process of structure formation. Galaxy mergers play a pivotal role in the evolution of galaxies by disturbing the kinematics of the stellar and gaseous components, which induces morphology transformation, size growth, star formation, and active galactic nuclei (AGNs) activities~\citep[e.g.][]{kauffmann00,springel05a,daddi05,trujillo06,croton06,schawinski06,cox08,naab09,hopkins10a,vandokkum10,oser12,ji14,kaviraj14b,kaviraj14a,kaviraj14c}. Therefore, galaxies, especially massive ones, are the end products of structure formation in the $\Lambda$CDM cosmology.

Many puzzles in the evolution of galaxies are still waiting to be addressed, and there is no clear consensus even on seemingly simple issues of the assembly history of stellar masses in galaxies in a quantitative aspect. Much effort has been made to systematically investigate these issues in a cosmological context. Semi-analytic models (SAMs) for galaxy formation and evolution have been used to explore the formation and assembly history of stellar components of galaxies in  cosmological volumes by taking advantage of their computing efficiency. Empirical downsizing effects~\citep[e.g.][]{cowie96,glazebrook04,cimatti04} are reasonably reproduced in SAMs based on the $\Lambda$CDM cosmology in which galaxies are born earlier on shorter timescales in denser environments and larger ones are built up through gradual mergers.~\citep[e.g.][]{delucia06,delucia07,guo08,jimenez11,lee13}.~\citet{delucia07} demonstrated that brightest cluster galaxies (BCGs) can have very high fractions of ex situ components ($>80\%$) in terms of total stellar mass.~\citet{jimenez11} examined the assembly history of model galaxies in two clusters, and found that BCGs acquire $35\%$ of their final mass via mergers, and the fractions monotonically decrease with increasing absolute magnitudes of galaxies.~\citet[][hereafter LY13]{lee13} provide a quantitative prediction of the ex situ fractions of a complete set of galaxies in a cosmological volume as a function of the final stellar mass: 20\%, 40\%, and 70\% for $\log M_{*}/M_{\odot}\sim$10.75, 11.25, and 11.75 galaxies at $z=0$, respectively. According to them, in the main progenitors of local massive galaxies ($\log M_{*}/M_{\odot}>11.5$ at $z=0$) galaxy mergers become the leading channel in mass growth at $z\sim2$. For comparison, there is no such transition in smaller galaxies ($\log M_{*}/M_{\odot}<11.0$ at $z=0$).

Rapid advances in computing power have enabled us to carry out hydrodynamic simulations with reliable resolution for numerous haloes in cosmological volumes.~\citet{oser10} investigated the stellar assembly history of massive galaxies by using zoom-in simulations of 39 haloes. They showed that only $\sim20\%$ of the stars in massive galaxies ($\log M_{*}/M_{\odot}\gtrsim11.5$ at $z=0$) are formed in situ, and the rest fall into the galaxies via mergers. The galaxies in the zoomed simulations performed by~\citet{lackner12} have ex situ fractions $\sim1/3$ of those in~\citet{oser10} in similarly massive galaxies. The simulations in the two studies were run without AGN feedback. The effect of AGN feedback on stellar mass growth was investigated using zoomed simulations by~\citet{dubois13}. AGN feedback effectively suppresses the in situ star formation in massive galaxies, resulting in a 30\% higher ex situ fraction than in non-AGN cases. In their zoomed simulations,~\citet{hirschmann15} showed that galactic winds play a role similar to the AGN feedback in mass growth.

Hydrodynamic simulations for entire cosmological volumes with moderate resolution have become available recently~\citep[e.g.][]{vogelsberger14,schaye15}.~\citet{rodriguez-gomez16} examined the contribution of mergers to the galaxy mass assembly history by using a complete set of galaxies identified in a cosmological volume of the Illustris simulation~\citep{vogelsberger14,genel14}. The ex situ fractions of the galaxies in the volume are in good agreement with \citetalias{lee13}, i.e., $\sim20\%$ for $\log M_{*}/M_{\odot}\sim11$ and $\gtrsim70\%$ for $\log M_{*}/M_{\odot}\sim12$ at $z=0$. They also found that the ex situ components are less concentrated than in situ components and half of them flow into galaxies via major mergers in the overall mass range ($\log M_{*}/M_{\odot}>9$ at $z=0$).

The aforementioned studies looked into the histories of galaxy assembly primarily as functions of stellar mass. 
However, the properties of galaxies are governed by halo evolution in the $\Lambda$CDM cosmology, as implied by the strong correlation between the stellar and halo masses~\citep[e.g][]{moster10,behroozi10,behroozi13}. Furthermore, recent deep imaging observations revealed that a considerable fraction ($\sim40\%$) of bright ($M_{r}<20$) early-type galaxies have post-merger signatures in both isolated and dense environments~\citep{vandokkum05,sheen12}. Post-merger signatures in satellite galaxies were expected to be rare due to the high peculiar velocities in dense environments.~\citet{yi13} argued that the post merger features of non-central galaxies in dense environments may have been pre-processed before becoming satellites. These studies point out that the galaxy assembly history should be traced along with the evolution of environments. 

The number of neighboring galaxies is widely used to quantify the environments around galaxies in empirical studies, and this is closely connected to the host halo mass~\citep[see][and references therein]{muldrew12}. Thus, host halo mass can be used as a reasonable proxy of the environments of galaxies. Large cosmological volume simulations are needed to cover a variety of environments. \citetalias{lee13} investigated the origin of stellar components only as a function of stellar mass without separating them into centrals and satellites. The size of the cosmological volume \citetalias{lee13} used was $99.4$Mpc on a side with $512^3$ collisionless particles, and less than 30 cluster-scale haloes ($\log M_{200}/M_{\odot}>14$) being found in the volume. Up-to-date hydrodynamic cosmological simulations cover a scale of volumes similar to \citetalias{lee13} , which are not large enough to harvest many cluster-scale haloes. Therefore, semi-analytic approaches are still effective for investigating galaxy evolution in larger cosmological volumes. As a follow-up of \citetalias{lee13}, this study aims to separately scrutinize the mass assembly history of central and satellite galaxies as a function of halo and stellar mass using our own SAM.

\section{Model}
We use \ysam \citepalias{lee13} in this investigation and provide a summary of it in this section. 

\subsection{Halo catalogue}

To obtain a sufficient number of haloes in $\log M_{200}/M_{\odot}\sim 10-15$, we performed a cosmological volume simulation with $1024^3$ collisionless particles in a 284Mpc ($200h^{-1}$Mpc) periodic cube using the cosmological simulation code \gadgett~\citep{springel05b}. The initial condition of the simulation was generated using \mpgrafic~\citep{prunet13}, a parallel version of \grafic~\citep{bertschinger95}. We adopted a set of cosmological parameters derived from the seven-year Wilkinson Microwave Anisotropy Probe observations by~\citet{komatsu11},  $\Omega_{\rm m}=0.272$, $\Omega_{\Lambda}=0.728$, and $h=0.704$. A total of 125 snapshots were printed out from the volume run, and \subfind~\citep{springel01} was used to search for sub-structures in the friends-of-friends groups of the snapshots. The final halo catalogue consists of 118 time steps from $z=15.8$ to $z=0$.

\begin{table*}
\centering
\begin{threeparttable}
 
  \caption{List of free parameters tuned for the model calibration}

  \begin{tabular}{lllll}
  \hline
   Parameter &  Description & Fiducial value  & Tuning range \\
 \hline
$f_{\rm scatter}\tnote{1}$	& Fraction of stellar mass in satellites scattered by mergers & 0.4   & 0.2-0.5\\ 
$\alpha_{\rm SF}\tnote{2}$			& Star formation efficiency of cold gas disks&  0.075  & 0.02-0.10\\
$\epsilon^{\rm SN}_{0}\tnote{3}$ 			& Efficiency of SN feedback &  1.5 & 0.5-2 \\
$\alpha_{\rm SN}\tnote{3}$ 		& SN feedback power-law slope &  2.9 & 2-3 \\
$V_{\rm SN}\tnote{3}$             & Characteristic velocity of SN feedback (km s$^{-1}$)  &  250   &150-250\\
$f_{\rm BH}\tnote{4}$ 		& Cold gas accretion efficiency onto SMBHs induced by mergers & 0.035  & 0.01-0.04\\
$\kappa_{\rm AGN}\tnote{5}$		& Efficiency of radio mode AGN feedback  &  $3.5\times10^{-5}$ & $10^{-5}-10^{-4}$\\
$\alpha_{\rm AGN}\tnote{5}$ 		& Radio mode AGN feedback power-law slope &  2.3 & 2-3 \\
$V_{\rm AGN}\tnote{5}$             & Characteristic velocity of radio mode AGN feedback (km s$^{-1}$)  &  150   &150-250\\

\hline
\end{tabular}
\begin{tablenotes}
\small
\item Note: $^{1}$~\citet{murante04}. $^{2}$ Eq. 16 in~\citet{croton06}. Since this prescription allows cold gas disks to form stars when their surface densities exceed critical values, this efficiency is higher than that (0.02) of the original prescription of \ysam\ (Eq. 5 in \citetalias{lee13}) in which star formation rates are in proportion to the total amount of cold gas. $^{3}$ Eq. 12 in~\citet{somerville08}. The characteristic velocity fixed at 200 km s$^{-1}$ in the equation was adjusted in this study. $^{4}$ Eq. 2 in~\citet{kauffmann00}. $^{5}$ Eq. 10 in~\citet{croton06}. The characteristic velocity and the power fixed at 200 km s$^{-1}$ and 3 in the equation were adjusted in this study.

\end{tablenotes}
\label{parameters}
\end{threeparttable}
\end{table*}

\subsection{Halo merger trees}

Halo merger trees are the essential backbones of SAMs, and are composed of single or multiple branches. In this study, the branch linking the most massive progenitor among all the progenitors of a halo at each time step is defined as the main branch of a halo merger tree~\citep[see][]{delucia07}. These trees are constructed from the halo catalogue using the tree building code \ysamtm~\citep{jung14}, which traces the most likely descendant or progenitor of a halo by comparing the identifications of particles bound to the haloes in two snapshots. \ysamtm\ has been updated to provide the number fraction of bound particles exchanged between haloes. We assume that the same fraction of hot gas and diffuse stellar components are transferred along with the collisionless particle exchanges between haloes.

\subsection{Semi-analytic model}

\ysam\ allows a halo to form a galaxy in its central region~\citep{white78}. In principle, a halo has one galaxy regardless of whether it is a host or sub. Satellite galaxies are treated as the centrals of subhaloes. \ysam\ performs tree cleaning and repairing processes before planting galaxies onto raw halo merger trees that possibly have problematic branches~\citep{muldrew11,elahi13,onions13,srisawat13}. Halo finding codes sometimes fail to identify haloes embedded in dense environments or close to the mass resolution limit, which eventually results in fragmented trees. \ysam\ removes any host halo branches that disappear without descendants before $z=0$ and subhalo branches that are identified to be newly formed with no progenitor. \ysam\ analytically calculates the orbits and mass of the subhaloes that merge into host haloes in the raw merger trees before reaching the central regions of their hosts~\citep[see][]{lee13,lee14}. We adopt the prescriptions for gas cooling proposed by~\citet{white91}. The prescription for quiescent star formation was updated from the original prescription in \ysam. Star formation in a disk is permitted when the surface density of a cold gas disk is larger than a critical density~\citep{croton06}. We use the prescriptions for supernova feedback and merger-induced starbursts proposed by~\citet{somerville08}. Quasar and radio mode feedback is implemented into \ysam\ as proposed by~\citet{kauffmann00} and~\citet{croton06}. We also trace the mass loss and chemical enrichment history of individual stellar populations in galaxies. Further details can be found in \citetalias{lee13}.

\begin{figure*}
\centering 
\includegraphics[width=0.8\textwidth]{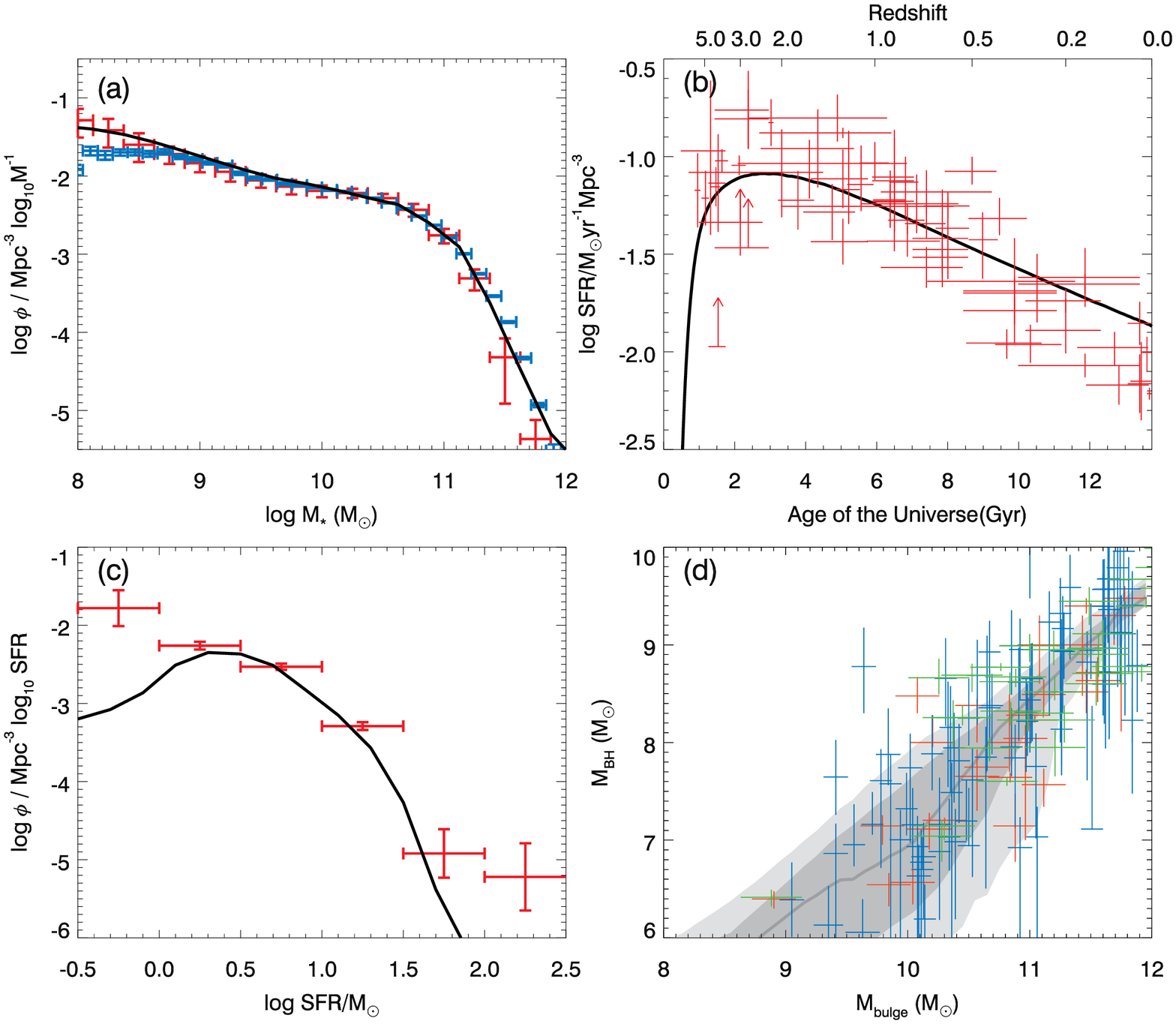}
\caption{Fitting of the fiducial model in this study. (a) Galaxy stellar mass function at $z\sim0$. The blue symbols mark the empirical GSMF derived by~\citet{panter07}. The red symbols are the composite GSMF of three different empirical GSMFs~\citep{baldry08,li09,baldry12}. The model MF is indicated by the black solid line. (b) Global star formation density evolution. The red crosses are the empirical data~\citep{panter07} and the black solid line displays the model evolution. (c) Star formation rate function at $z\sim0.15$ for $M_{*}>10^{10}M_{\odot}$. The red symbols present the empirical data~\citep{gruppioni15}. (d) The supermassive black hole-bulge stellar mass relation. The grey solid line, dark shade, and light shading indicate the mean, $16-84$th percentile distribution, and $2.5-97.5$th percentile distribution of $P(M_{\rm BH}|M_{\rm bulge})$, respectively. The red, blue, and green data points with error bars are from~\citet{haring04},~\citet{kormendy13}, and~\citet{mcconnell13}, respectively.}
\label{fits}
\end{figure*}

\subsection{Model calibrations}

\ysam\ was calibrated for the cosmological volume described in \S2.1 by tuning the set of free parameters listed in Table 1. More parameters are used in the prescriptions of \ysam\ \citepalias[see][and references therein]{lee13}. Most of them are, however, fixed and we have mainly adjusted the listed free parameters that regulate feedback and star formation efficiency. Figure~\ref{fits} shows the fitting of our fiducial model. The galaxy stellar mass function (GSMF) is adopted as the primary calibration point for \ysam. The panel (a) in Figure~\ref{fits} displays two empirical GSMFs along with our fiducial model at $z=0$. The GSMF marked by blue symbols comes from~\citet{panter07}, and the red symbols are a composite GSMF of~\citet{baldry08},~\citet{li09}, and~\citet{baldry12}. The two empirical GSMFs are in good agreement overall, even though that of~\citet{panter07} has a slightly higher massive end ($M_{*}>10^{11}M_{\odot}$). Our fiducial model is located in between these two GSMFs. The panel (b) in Figure~\ref{fits} shows the evolution of the global star formation density (GSFD) over the cosmic time. The empirical GSFD that was compiled and modified by~\citet{panter07} is marked by red crosses. In panel (c), one can see the star formation rate functions (SFRFs) of the empirical data~\citep{gruppioni15} and our fiducial model at $z\sim0.15$ for $M_{*}>10^{10}M_{\odot}$. The GSFD and SFRF are not calibration points in our model, but are used to cross-check whether our model reproduces reliable star formation histories.

A tight correlation between the mass of supermassive black holes (SMBHs) and bulge stellar mass has been discovered~\citep{magorrian98,marconi03,haring04,hu09,sani11}. This relation was used as the secondary calibration point in our model. Three different observations with error bars and $P(M_{\rm BH}|M_{\rm bulge})$ of the fiducial model are shown in panel (d) of Figure~\ref{fits}. The red, blue, and green crosses indicate the empirical $M_{\rm BH}-M_{\rm bulge}$ relation derived by~\citet{haring04},~\citet{kormendy13}, and~\citet{mcconnell13}, respectively. The $P(M_{\rm BH}|M_{\rm bulge})$ distribution of our model is within the observational scatter. A notable feature is that the distribution of the model at $\log M_{\rm bulge}/M_{\odot}>11$ is much narrower than that in the low mass range. This is because the main channel that builds up bulges at the massive end is galaxy mergers, which also feed the SMBHs by injecting gas into very central regions or by inducing SMBH mergers in our model. Dry mergers  strengthen the relationship in the later stages of the evolution of massive galaxies. On the other hand, disk instability becomes significant in the growth of (pseudo) bulges for less massive galaxies along with mergers. This results in a less tight $M_{\rm BH}-M_{\rm bulge}$ relation at $\log M_{\rm bulge}/M_{\odot}<11$, as suggested by the empirical studies~\citep[see][]{kormendy13b}. This model is used as a fiducial model in this study.

\begin{figure*}
\centering 
\includegraphics[width=0.85\textwidth]{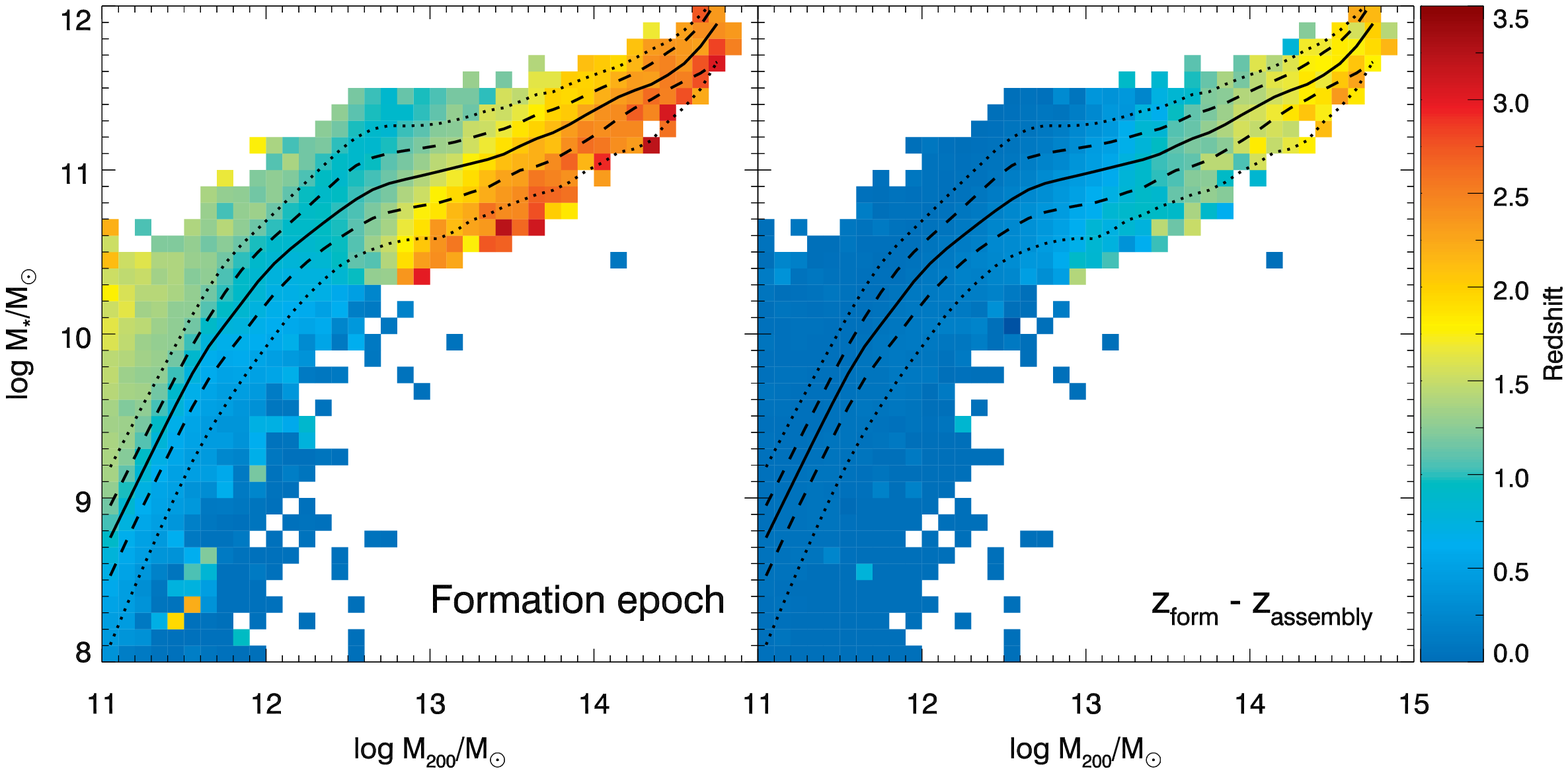}
\caption{Left: Distribution of formation redshifts by which half of the stellar mass in central galaxies at $z=0$ has been formed in the stellar-to-halo mass plane. Right: Redshift differences between the formation and assembly epochs when half of the stellar mass is located in the final centrals. Bluer when $z_{\rm form}\sim z_{\rm assembly}$. The black solid, dashed, and dotted lines mark the median, 16-84th percentile distribution, and 2.5-97.5th percentile distribution of $P(M_{*}|M_{\rm 200})$, respectively.}
\label{age}
\end{figure*}

\section{Results}

All of the galaxies in \ysam\ are born as the central galaxy of each halo. Thus, the final galaxy of a halo merger tree composed of {\it N} branches evolves with a maximum of {\it N}-1 galaxy mergers. The galaxies that grow along the main branches of the halo merger trees are called the main or direct progenitors of the final galaxies. The stars formed in the main progenitors are classified as in situ components, and those that are formed outside and then come into the main branches are labelled as ex situ components. In this definition, the stellar mass $M_{*}$ of the galaxy is the sum of the mass of the two components, i.e., $M_{*}=M_{\rm in\, situ}+M_{\rm ex\, situ}$, following a conventional nomenclature~\citep[e.g.][]{oser10,lee13,rodriguez-gomez16}. We simply assume that the formation of in situ components is a self-assembly process. A host halo that habors {\it N} subhaloes has one central galaxy and a maximum of {\it N} satellite galaxies.

\subsection{Formation and assembly history of stellar mass}

We quantify the star formation history of a galaxy by finding the redshifts $z_{\rm form}$ by which half of the stellar populations found in the final galaxies have been formed. The assembly time of a galaxy is defined as the epoch $z_{\rm assembly}$ at which the stellar mass of the main progenitor reaches half of the final mass. The equivalence $z_{\rm form}=z_{\rm assembly}$ is made only when galaxies evolve along single-branch merger trees. Figure~\ref{age} shows the distributions of $z_{\rm form}$ and $z_{\rm form}-z_{\rm assembly}$ of the central galaxies in the stellar-to-halo mass plane at $z=0$. The solid, dashed, and dotted lines mark the median, 16-84th percentile distribution, and 2.5-97.5th percentile distribution of $P(M_{*}|M_{\rm 200})$. The downsizing trend is clearly shown as a function of the halo mass. This figure implies that the empirical downsizing trend in which more massive galaxies have older stellar ages originates from the dependence of $z_{\rm form}$ on the halo mass and stellar-to-halo mass relation. The formation time of very massive galaxies ($\log M_{*}/M_{\odot}>11.5$ at $z=0$) is found at $z_{\rm form}>2$, in agreement with~\citet{thomas10} who derived the half mass formation time of early type galaxies using the Sloan Digital Sky Survey.


\citetalias{lee13} found an upsizing trend in the assembly time of stellar mass, which is directly opposite to the downsizing nature of formation time. This result is consistent with recent observations.~\citet{bundy09} found from the GOODS field that the pair fraction of red massive spheroidals $(\log M_{*}/M_{\odot} > 11)$ is higher than that of smaller galaxies at $z\sim1$. They expected that the pairs would eventually merge with each other, and form massive early types.~\citet{matsuoka10} showed that the number density of massive galaxies $(\log M_{*}/M_{\odot}\sim11.5-12)$ increases faster than that of less massive galaxies $(\log M_{*}/M_{\odot}\sim11-11.5)$ for $0<z<1$. The majority of the massive galaxies appear to be quenched. Consequently, mergers are their most likely mass growth channel during the period of time. In our fiducial model, the centrals of more massive haloes have younger assembly times, which results in the increase of $z_{\rm form}-z_{\rm assembly}$ with increasing halo mass (the right panel of Figure~\ref{age}). In summary, more massive galaxies are older in terms of formation ages but relatively younger in terms of assembly ages~\citep{delucia06,delucia07,guo08,jimenez11,lee13,lee14}, and this trend originates from the halo mass assembly.

\subsection{Origin of stellar components}

\subsubsection{Ex situ fractions in the stellar-to-halo mass plane}

\begin{figure*}
\centering 
\includegraphics[width=0.85\textwidth]{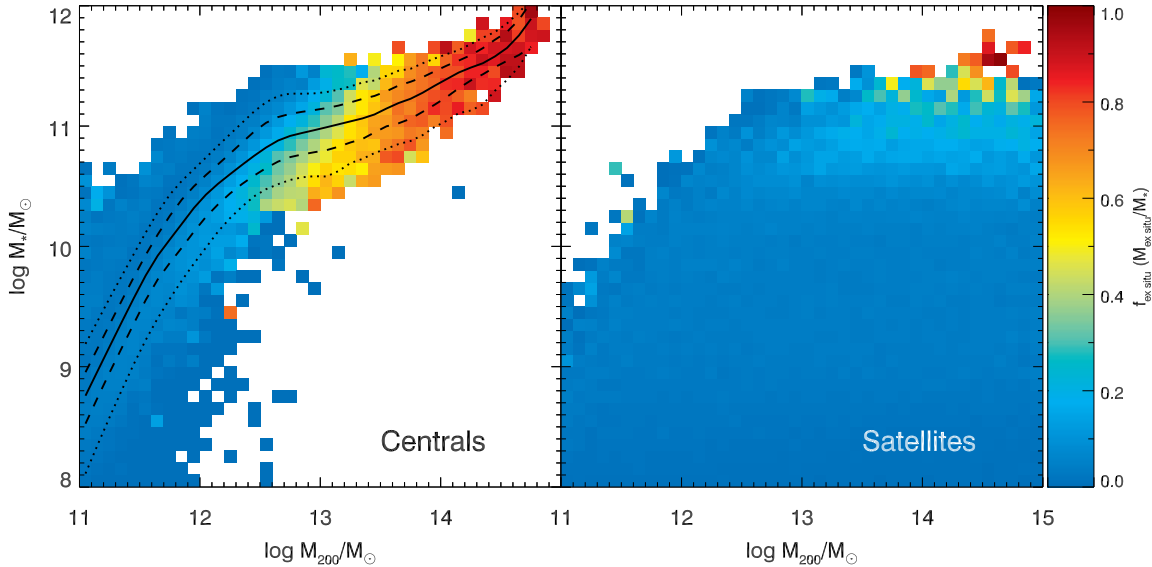}
\caption{The fractions of ex situ components in the stellar-to-host halo mass plane. Left and right panels are for central and satellite galaxies, respectively. The $M_{200}$ of the satellites is their host halo mass.}
\label{f_acc}
\end{figure*}

We quantified the overall contribution of galaxy mergers to stellar mass growth by measuring the fraction of ex situ components in galaxies, i.e., $f_{\rm ex\, situ}=M_{\rm ex\, situ}/M_{*}$. Figure~\ref{f_acc} shows the $f_{\rm ex\, situ}$ distribution of central and satellite galaxies in the stellar-to-host halo mass plane at $z=0$. The $f_{\rm ex\, situ}$ distribution of the centrals has a strong correlation with $M_{200}$, but its dependency on $M_{*}$ is not clear. This result suggests once again that the increase in $f_{\rm ex\, situ}$ with increasing $M_{*}$ demonstrated by previous studies~\citep[e.g.][]{oser10,lackner12,lee13,rodriguez-gomez16} may be a derivative of the $f_{\rm ex\, situ}-M_{200}$ and the stellar to halo mass relations. 

The stellar-to-halo mass relation of satellites is broken up as haloes are stripped during orbital motion in dense environments~\citep[e.g.][]{boylan-kolchin08}. Therefore, the $f_{\rm ex\, situ}$ distribution of satellites in the right panel of Figure~\ref{f_acc} is plotted as a function of their host halo mass. As the stellar-to-halo mass relation is shuffled, no correlation was found between the $f_{\rm ex\, situ}$ distribution of satellites and their host halo mass. However, the correlation between $f_{\rm ex\, situ}$ and the mass of satellites appears to be clearer than that of centrals. When they are in the same halo, satellite galaxies are generally less massive than centrals  mainly because of low mass growth rates after becoming satellites. Tidal or ram pressure stripping  effectively removes gas reservoirs in satellites, eventually quenching star formation activities~\citep{gunn72,abadi99,quilis00,chung07,tonnesen09,yagi10,kimm11,steinhauser12}. In addition, there is little opportunity  to increase their mass via mergers.

\begin{figure}
\centering 
\includegraphics[width=0.5\textwidth]{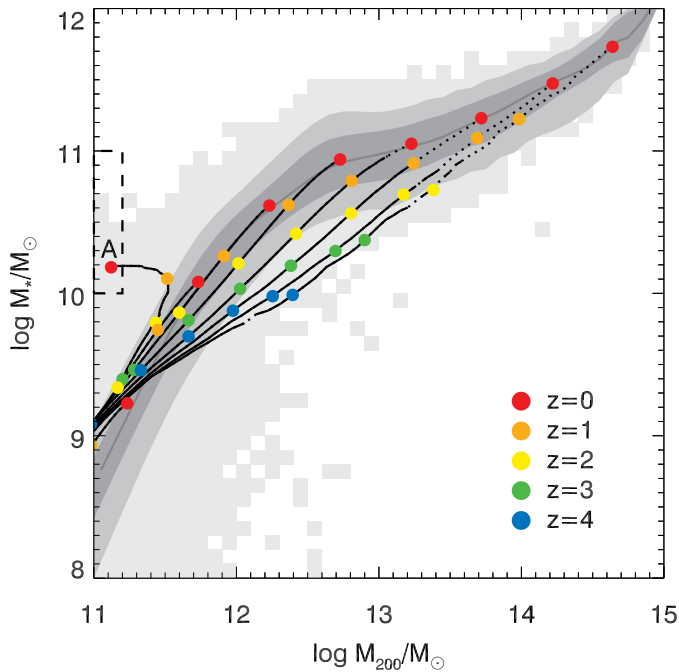}
\caption{Mean halo and stellar mass evolution histories of galaxies on the stellar-to-halo mass plane. The grey solid line, dark shading, and light shading show the median, $16-84$th percentile distribution, and $2.5-97.5$th percentile distribution of $P(M_{*}|M_{\rm 200})$ at $z=0$. The filled red circle indicated by \textbf{A} is the averaged mass of the galaxies in the mass range marked by the dashed box. The other filled red circles indicate the averaged galaxies binned by every 0.5 dex of halo mass from $\log M_{200}/M_{\odot}=11$ at $z=0$. The black solid and dotted lines show mass evolution tracks of the main progenitors of the averaged galaxies and the colored circles point out redshifts. The epochs when in situ star formation or mergers lead mass growth are marked by solid or dotted lines, respectively.}
\label{mass_evolution}
\end{figure}

We trace the mass growth histories of central galaxies in the stellar-to-halo mass plane. Figure~\ref{mass_evolution} shows the mean mass evolution tracks of the main progenitors of galaxies binned by final mass. Since we defined $M_{*}=M_{\rm in\, situ}+M_{\rm ex\, situ}$, the mass growth rate of a galaxy at an epoch is $\dot{M}_{*}=\dot{M}_{\rm in\, situ}+\dot{M}_{\rm ex\, situ}$. The black solid lines represent $\dot{M}_{\rm in\, situ}\geq\dot{M}_{\rm ex\, situ}$ phase and the dotted lines represent the opposite. The color code of the filled circles indicates redshifts. It can be seen that halo mass regulates which channel is dominant in mass growth. The central galaxies of the most massive haloes ($\log M_{200}/M_{\odot} (z=0)\sim14.7$) enter the merger-dominant phase at $z\sim2$ while those at $\log M_{200}/M_{\odot} (z=0)\sim13.2$ only reach this phase until $z=0$. This plot shows that galaxies migrate to the merger-dominant phase when their host halo mass begins to exceed $\sim10^{13}M_{\odot}$. As mentioned above, more mergers in the centrals of larger haloes cause earlier quenching and higher ex situ fractions. Galaxies residing in haloes below $\log M_{200}/M_{\odot}\sim13$ increase their stellar mass in directly proportion to the increase in their halo mass. For example, central galaxies at $\log M_{200}/M_{\odot} (z=0) \sim12.7$ experience 1.4 dex of halo mass growth between $z=4-0$ and their stellar mass increases to almost the same scale during the same period of time. On the other hand, halo mass growth does not accompany the same rate of stellar mass growth in the most massive group. This is because star formation is suppressed by strong feedback and not all of the stellar mass is concentrated in the central regions. Stars in massive haloes are located in satellite galaxies or intra-cluster~\citep{feldmeier02,gonzalez05} as well as in the central galaxies of massive haloes.

\begin{figure*}
\centering 
\includegraphics[width=0.85\textwidth]{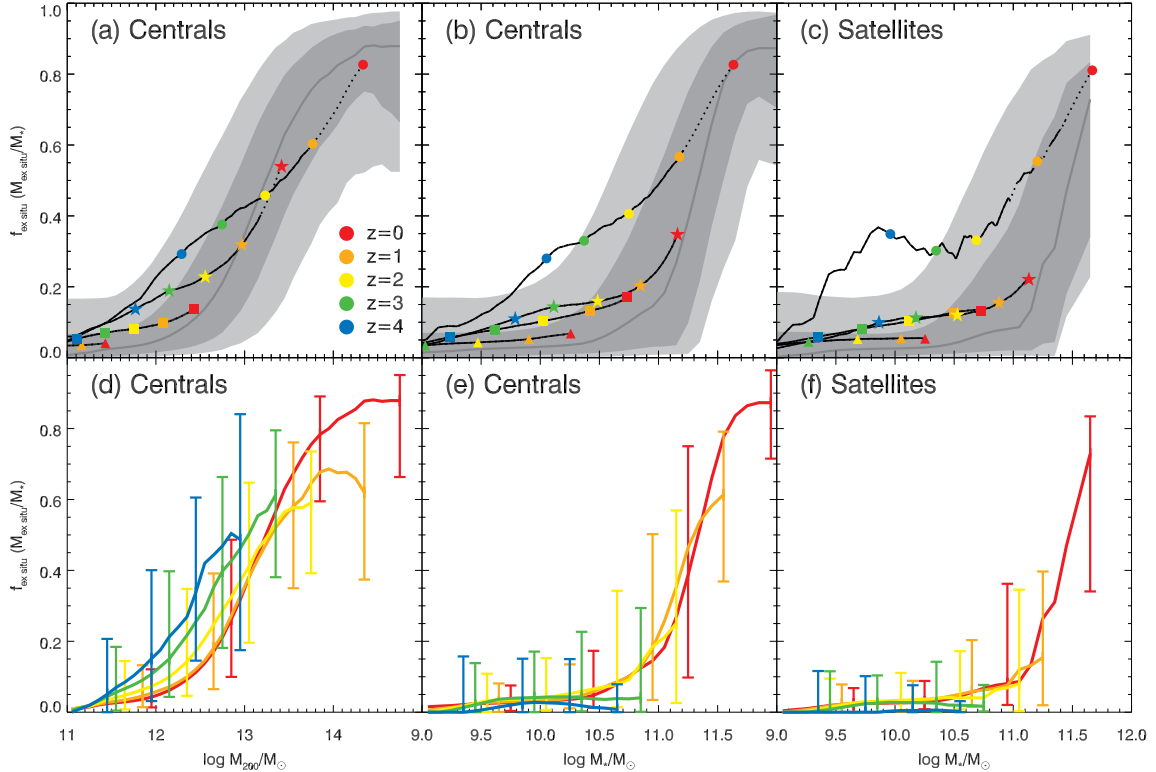}
\caption{Upper: Marginal $f_{\rm ex\, situ}$ distribution in terms of halo or stellar mass. The grey solid lines, dark shading, and light shading show the median, $16-84$th percentile distribution, and $2.5-97.5$th percentile distribution of $P(f_{\rm ex\, situ}|M)$ at $z=0$. Each black line with the same symbol displays the mean $f_{\rm ex\, situ}-M$ evolution of the main progenitors of galaxies binned by the final halo or stellar mass. The epochs when in situ star formation and mergers lead mass growth are marked by solid and dotted lines, respectively. The colored symbols indicate the five different redshifts, according to the color code in panel (a). The most massive, second most massive, and third most massive groups are indicated by red circles, stars, and squares, respectively, in panels (b) and (c) and correspond to the massive galaxy groups in \citetalias{lee13}. Bottom panels: Redshift evolution of the $f_{\rm ex\, situ}-M$ relation. The solid lines with the error bars mark the median and 16-84th percentile distribution of $f_{\rm ex\, situ}$. The color code shown in Panel (a) is also used to indicate the redshifts. Mass bins including at least five galaxies are plotted in all the panels.}
\label{f_acc2}
\end{figure*}

The peak of the stellar-to-halo mass fraction is found at $\log M_{200}/M_{\odot}\sim12$ as $\log M_{*}/M_{200}\sim-1.5$~\citep{moster10,behroozi10,behroozi13}. However, the galaxies in the dashed box have notably large fractions. Their number fraction is actually negligible, far outside the 97.5th percentile distribution of $P(M_{*}|M_{200})$. Only 87 galaxies are found in the box out of the $100,000$ plus galaxies at $\log M_{200}/M_{\odot}=11.0-11.2$ at $z=0$. We look into how the galaxies gain that particularly large $M_{*}/M_{200}$ ratio. As the averaged evolution track of the galaxies in the box marked by \textbf{A} indicates, they undergo severe halo stripping at $z<1$. They barely increase their stellar mass during the period of time because of gas loss. Besides, the degree of halo stripping is not harsh enough to significantly strip stellar components in central regions~\citep{smith16}. All of the central galaxies are typically located in dense environments where tidal force exerted by neighbouring massive haloes disturbs small systems.

\subsubsection{Ex situ fractions at a given stellar and halo mass}

The marginal distribution of $f_{\rm ex\, situ}$ in terms of $M_{200}$ or $M_{*}$ is plotted in Figure~\ref{f_acc2}. The grey shading displays the percentile distribution of $P(f_{\rm ex\, situ}|M)$. Panel (a) demonstrates the gradual increase of $f_{\rm ex\, situ}$ of the centrals with increasing $M_{200}$. However, in panel (b) the $f_{\rm ex\, situ}$ of the central galaxies stays below 0.1 at $\log M_{*}/M_{\odot}<10.5$ and rises sharply in $\log M_{*}/M_{\odot}>11$. This is because galaxies with $\log M_{*}/M_{\odot}\sim11$ at $z=0$ are hosted by haloes in the wide mass range of $\log M_{200}/M_{\odot}\sim12-14.5$. The relation between the stellar and halo masses causes the large dispersion of $f_{\rm ex\, situ}$ around $\log M_{*}/M_{\odot}=11$. The satellites show a slightly lower $f_{\rm ex\, situ}-M_{*}$ relation than centrals, but the overall trend is similar, as shown in panel (c). Specifically, the $f_{\rm ex\, situ}$ of the most massive satellites is comparable to that of the centrals. This suggests that mergers are essential to form massive galaxies. Such massive galaxies were centrals until recently, and have only just became satellites.

The black solid and dotted lines with colored symbols in the upper panels of Figure~\ref{f_acc2} show the averaged evolution tracks of the main progenitors in the $f_{\rm ex\, situ}-M$ planes. The color code and line styles are the same as those in Figure~\ref{mass_evolution}. The galaxies are binned by 1 dex from $\log M_{200}/M_{\odot}=12$ or 0.5 dex from $\log M_{*}/M_{\odot}=10$ at $z=0$. The main progenitors of the galaxies in the different mass bins are marked by different symbols. Panel (a) illustrates that galaxies that become merger-dominant before $z=0$ are finally hosted by haloes of $\log M_{200}/M_{\odot}>13$. In panels (b) and (c), the transition takes place only in the most massive groups of both centrals and satellites. Stochastic effects cause the uneven tracks of the most massive satellites in panel (c). 

 \citetalias{lee13} provided a quantitative prediction of the mean $f_{\rm ex\, situ}$ as a function of the final stellar mass without separating galaxies into central and satellites: 20\%, 40\%, and 70\% for the galaxies in $\log M_{*}/M_{\odot}=$10.5-11, 11-11.5, and 11.5-12 at $z=0$, respectively. The red filled circles, stars, and squares, in panels (b) and (c) of Figure~\ref{f_acc2} indicate the mean $f_{\rm ex\, situ}$ of the three groups binned by the final stellar mass. Because the majority of the galaxies in the three groups are centrals, the $f_{\rm ex\, situ}-M_{*}$ relation of the centrals is comparable to that of \citetalias{lee13}. The $f_{\rm ex\, situ}-M_{*}$ relation and the large dispersion at $\log M_{*}/M_{\odot}\sim11$ are in good agreement with \citetalias{lee13} and~\citet{rodriguez-gomez16}.

The bottom panels of Figure~\ref{f_acc2} demonstrate the evolution of the $f_{\rm ex\, situ}-M$ relation at $z\sim0-4$. The different redshifts are indicated by the color codes used in panel (a). At a given halo mass, the central galaxies have a higher $f_{\rm ex\, situ}$ at higher redshifts. This is because the haloes that show up at higher redshifts are located in relatively denser environments than those having the same mass at lower redshifts. Therefore, low mass haloes at $z\sim4$ likely have growth histories radically different from local low mass haloes. On the other hand, high redshift galaxies have a low median $f_{\rm ex\, situ}$ in panels (e) and (f). The most massive haloes are always rare and their centrals have stellar masses with larger dispersions at higher redshifts. In other words, the stellar-to-halo mass relation is less tight at higher redshifts. The ex situ fraction begins to rapidly increase at $\log M_{*}/M_{\odot}=11$, which implies that galaxy mergers are essential to build up galaxies above $\log M_{*}/M_{\odot}=11$. On the contrary, galaxies similar in mass to the Milky Way ($\log M_{*}/M_{\odot}\sim10.5$) would increase their stellar mass mostly from in situ star formation, regardless of whether they are centrals or satellites.

\subsection{Evolution history of the two mass growth channels}

\subsubsection{Specific mass growth histories of the main progenitors}
The previous sections show that the leading mass growth channels change over time, depending on the halo or stellar mass. In order to examine the time evolution of the contribution of mergers and in situ star formation to stellar mass growth, we measure the specific stellar mass accretion rate (SSAR) $\dot{M}_{\rm ex\, situ}/M_{*}$ in the same way that the specific star formation rate (SSFR) is defined. In this study, accretion indicates the infall of stellar components into galaxies only via mergers.

\begin{figure}
\centering 
\includegraphics[width=0.45\textwidth]{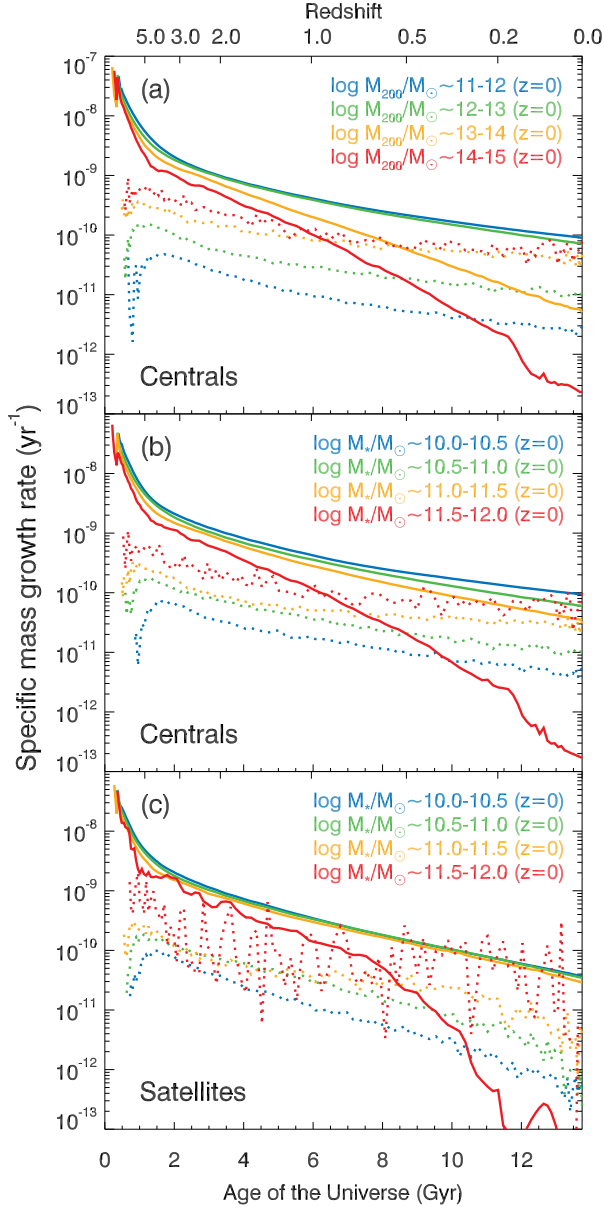}
\caption{Evolution of the mean specific stellar mass growth rate for the main progenitor galaxies binned by final halo or galaxy mass. The color code indicates the given mass ranges, and the solid and dashed lines indicate the specific star formation rates and specific stellar mass accretion rates via galaxy mergers, respectively.}
\label{ssfr_ssar0}
\end{figure}

Figure~\ref{ssfr_ssar0} shows the SSFRs and SSARs of the main progenitors of galaxies grouped by final mass and status. In all the panels, the SSFRs are lower in more massive galaxies, while the opposite is true for SSARs, which gently and consistently decrease. However, the SSFRs rapidly decline by $z\sim4$ after which the decay rates decelerate. In all the panels, the two channels decrease similarly in the third and fourth most massive groups at $z<3$. However, bigger groups experience a sharp decay in the SSFRs. As discussed in \S3.2.1, galaxies enter a merger-dominant phase when their host halo mass increases above $\log M_{200}/M_{\odot}\sim13$ in our model. Central galaxies in larger haloes are effectively quenched by more frequent AGN activities and acquire more stellar mass via more mergers. These two phenomena give rise to a transition of dominant mass growth channels. Mergers become the primary process in mass growth until z=0 for the main progenitors of central galaxies with haloes of $\log M_{200}/M_{\odot}>13$.

\begin{figure}
\centering 
\includegraphics[width=0.5\textwidth]{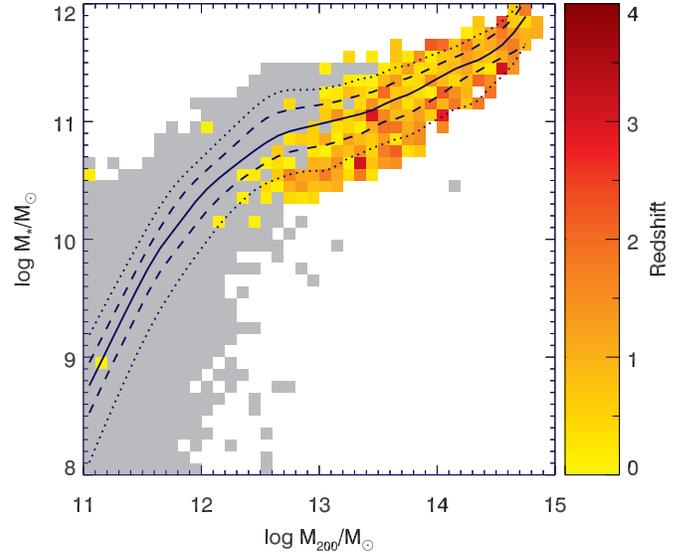}
\caption{Transition redshifts at which galaxy mergers becomes the dominant channel in mass growth. The grey shade mark galaxies that do not experience the phase transition by $z=0$.}
\label{z_acc}
\end{figure}

The main progenitors of the most massive centrals in panel (b) enter the merger-dominant phase at $z\sim1$. The SSAR of the second most massive group, however, does not meet its SSFR at all. Galaxies in this group are finally hosted by haloes in a wide mass range of $\log M_{200}/M_{\odot}\sim12-14.5$. Some of the galaxies residing in haloes above the group scales ($\log M_{200}/M_{\odot}>13$) finally settle into the merger-dominant phase, as shown in panel (a). However, the majority of the second massive group are the centrals of the smaller haloes ($12<\log M_{200}/M_{\odot}<13$). Thus, the main progenitors of the second most massive galaxies marginally stay in the SSAR$\lesssim$SSFR phase by $z=0$. This result is consistent with~\cite{woo13}, who found a strong correlation between the quenched fractions of central galaxies and their host halo mass. AGN feedback and decreasing cooling efficiency naturally result in the trend found in \ysam.

The evolution trend of the specific mass growth rates of satellites in panel (c) appears to be almost the same as that of the centrals in panel (b) in early epochs. This is primarily because they all used to be centrals at high redshifts. As they gradually become satellites with decreasing redshifts, their specific growth rates are suppressed. Because the satellites barely merge with each other, their SSARs are lower than those of the centrals.   
Environmental effects, such as tidal and ram pressure stripping, suppress star formation activities in satellites by blowing away gas reservoirs. However, AGN feedback is inactive in satellites due to little gas accretion. Therefore, the most and second most massive satellites eventually have slightly lower SSFRs than those of the centrals. The main progenitors of the most massive satellites exhibit behavior similar to those of the most massive centrals. This is mostly because they only recently became satellites. However, the SSARs of the other groups do not even come close to their SSFRs.

The transition epochs of the two mass growth channels as a function of the final stellar mass are in good agreement with previous studies that were based on semi-analytic approaches~\citep[e.g.][]{lee13} and hydrodynamics~\citep[e.g.][]{oser10,rodriguez-gomez16}. In Figure~\ref{z_acc}, the redshift transitions between the two channels are plotted in the stellar-to-halo mass plane. The grey shading indicates the mass ranges of the galaxies that stay in the star-formation-dominant phase through $z=0$ on average in each mass bin. The transitions are found in galaxies that end up in haloes with $\log M_{200}/M_{\odot} \gtrsim 13$, as panel (a) of Figure~\ref{ssfr_ssar0} demonstrates. This suggests that the halo mass evolution plays a key role in the two-phase scenario of massive galaxy formation in which in situ star formation rapidly increases the galaxy stellar mass at early epochs and mergers gradually build it up by $z=0$~\citep{oser10}.

\begin{figure}
\centering 
\includegraphics[width=0.45\textwidth]{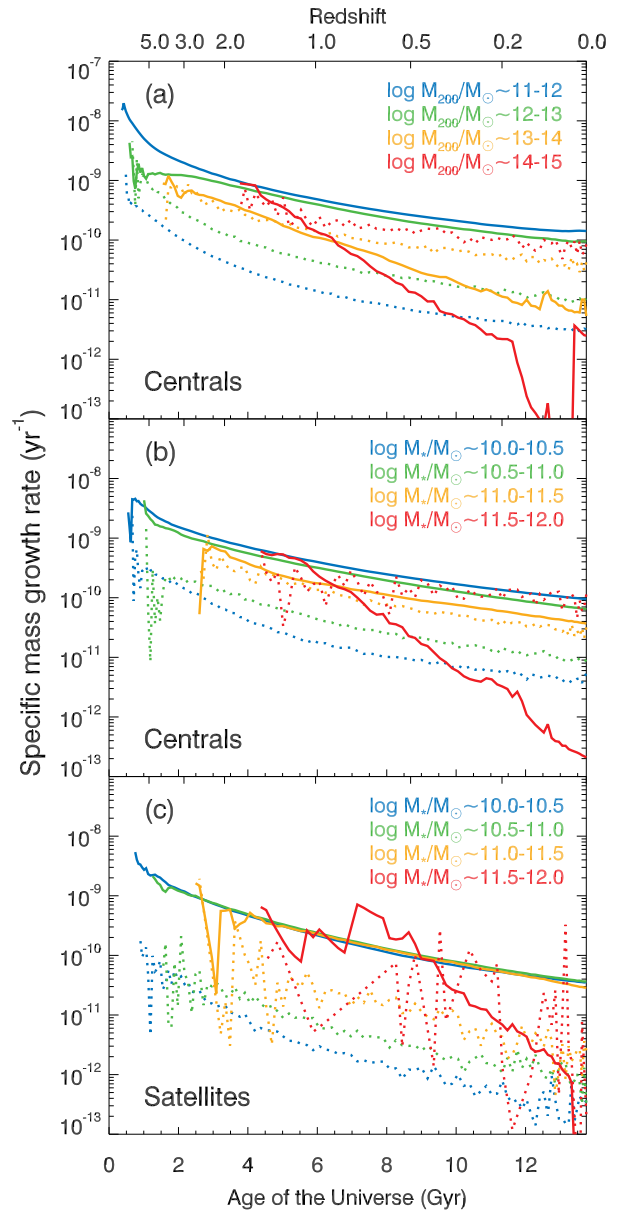}
\caption{Mean specific stellar mass growth rate of galaxies binned by the halo or stellar mass at each epoch. The color code and line styles are the same as those in Figure~\ref{ssfr_ssar0}.}
\label{ssfr_ssar1}
\end{figure}

\subsubsection{Specific mass growth rates at each epoch}

The specific mass growth rates of the main progenitors in the previous section indicate how the galaxies chosen at $z=0$ have evolved over time. Thus, Figure~\ref{ssfr_ssar0} is plotted based on a theoretical viewpoint. However, observations take a snapshot of the galaxies located at various redshifts . In that sense, Figure~\ref{ssfr_ssar1} displays the specific mass growth rates of galaxies binned by mass at each epoch. The more massive galaxies or the centrals of more massive haloes have higher SSARs and lower SSFRs. At a fixed stellar mass, the SSFRs decay by two orders of magnitude during cosmic time, as shown in previous studies~\citep{rodighiero10,karim11,sparre15,furlong15}. 

Cluster-scale haloes ($\log M_{200}/M_{\odot}>14$) and very massive galaxies ($\log M_{*}/M_{\odot}>11.5$) begin to appear at $z\approx2$ in the co-moving volume of this study. The SSFRs of the most massive groups are comparable to those of less massive groups for a while after emerging in all the panels.  The SSFRs are sufficiently high at the moment to increase the stellar mass by a factor of two within a billion years. In our model, galaxies in the most massive ranges at early epochs have been built up by very active star formation, along with galaxy mergers. Rapid halo mass growth leads to violent baryonic accretion into the central regions, eventually inducing very high star formation activities. However, the most massive groups quickly experience a rapid drop in the SSFRs in all the panels.

The central galaxies of the haloes in $\log M_{200}/M_{\odot}>13$ of panel (a) are in the merger-dominant phase most of the time after they appear in all the panels. The differences between the SSFRs and the SSARs increase in the massive halo groups as cold gas is depleted and cooling is further suppressed by feedback. In the middle panel, the two mass growth channels are almost equally important for the second massive group of centrals ($\log M_{*}/M_{\odot}\sim11.0-11.5$) all the time. In this panel, 0.5 dex more massive central galaxies have SSARs 0.5 dex higher at $z<0.5$. Consequently, a ten times larger stellar mass falls into 0.5 dex more massive galaxies via mergers. Star formation contributes an order of magnitude more to mass growth than mergers in $\log M_{*}/M_{\odot}<11$.

The specific mass growth rates of satellites (panel (c)) are always lower than those of centrals. Environmental effects cause overall low SSFRs. Since satellite galaxies are hardly involved in mergers with other satellites, their SSARs are at least an order of magnitude lower than those of the centrals. Therefore, mergers do not become primary mass growth channels in all groups. In our model, galaxy mergers between satellites mainly occur in sub groups that belong to host haloes that are not yet virialized. Once the sub groups are dynamically dissociated in dense environments, their member galaxies only fall into the central regions of their host haloes  in our model.

\section{Summary and Conclusion}
\citetalias{lee13} examined the assembly history of stellar components in galaxies as a function of final stellar mass using semi-analytic approaches. Here we expand \citetalias{lee13} in a cosmological context by investigating it in terms of extended parameter spaces using a larger cosmological volume simulation. The size and resolution of the volume were designed to cover a wide range of halo masses $\log M_{200}/M_{\odot}\sim10-15$. The fiducial model of this study was calibrated to fit a set of empirical data. We labelled the stars formed along the main branches of halo merger trees as in situ components and the rest of the stars falling into galaxies via mergers as ex situ components.

The $\Lambda$CDM cosmology predicts earlier formation of primordial structures in denser environments and their hierarchical assembly over time. In this framework, the formation and assembly of galaxies are expected to correlate with the evolution of halo environments. In our model, we found that centrals of more massive haloes have older formation times and higher ex situ fractions. The marginal distribution of the ex situ fractions at $z=0$ gradually increases with increasing halo mass. The ex situ components become the majority in the stellar mass of the central galaxies in $\log M_{200}/M_{\odot}>13$. However, the distribution of ex situ fractions sharply rises in terms of stellar mass with large dispersion in $\log M_{*}/M_{\odot}>11$. This is because the central galaxies at $\log M_{*}/M_{\odot}\sim11$ are hosted by a wide range of halo mass $\log M_{200}/M_{\odot}\sim12-14.5$. As a result, they have similar masses despite their diverse evolution tracks. 

Satellite galaxies have slightly lower ex situ fractions than centrals but the overall trend is similar. Like massive centrals, massive satellites acquired a considerable fraction of their stellar mass via mergers mainly when they were centrals and have only recently become satellites. The marginal distribution of the ex situ fractions evolves with decreasing redshifts. The centrals of the most massive haloes already reach $f_{\rm ex\, situ}\sim0.5$ at $z=4$. The ex situ fraction rapidly increases as the galaxy stellar mass begins to exceed $\log M_{*}/M_{\odot}\sim11$. This can be interpreted to suggest that galaxy mergers are essential for building up galaxies above $\log M_{*}/M_{\odot}\sim11$, whether they end up becoming centrals or satellites. 

We examined the time evolution of the specific star formation rates (SSFRs) and specific stellar mass accretion rates (SSARs). First, we traced the specific growth rates of the main progenitors of the galaxies grouped by final halo or stellar mass. The SSFRs of the main progenitors are always lower in more massive groups while the SSARs behave in the opposite manner. In very early epochs, the SSFRs far exceed the SSARs in all galaxies, but this dominance rapidly decreases over time. Furthermore, the SSFRs decay even faster for more massive galaxies, which results in a crossing of the two mass growth channels in some cases. The transition takes place in the main progenitors of the central galaxies that finally reside in the haloes of $\log M_{200}/M_{\odot}>13$. In our model, this is the mass range where the two phase scenario for massive early types~\citep{oser10} is valid. With the aforementioned results, this suggests that the correlation between the stellar mass and the empirical galaxy formation time~\citep[e.g.][]{cowie96}, and the fraction of ex situ components proposed by theoretical studies~\citep[e.g.][]{oser10,lackner12,lee13,rodriguez-gomez16} may be merely the projection of their intrinsic halo mass dependence upon the stellar-to-halo mass relation. When we looked into the SSAR and SSFR evolution of galaxies in an empirical sense, i.e. binning them by given mass ranges at each epoch with no use of progenitor-descendant relations, we found that mergers are a major channel for mass growth at all times in the centrals of $\log M_{*}/M_{\odot}>11$ or the centrals of haloes with $\log M_{200}/M_{\odot}>13$. However, in satellites, mergers are secondary or even negligible.   

This study displayed a strong correlation between the formation and assembly of galaxies and halo mass which only represents local environments. In large scales, however, weaker correlations have been found.~\citet{croton08} compared the empirical luminosity function of void galaxies with their SAM, concluding that large-scale environments do not significantly affect galaxy properties but halo mass assembly is a decisive factor in galaxy evolution.~\citet{jung14} also presented a similar result using \ysam\ that galaxy growth rates are largely insensitive to large-scale environments. So, in sum, the evolution of local halo environments plays a leading role in the formation and assembly of galaxy stellar mass. Central galaxies in dense environments grow with vigorous gas inflow at early epochs and the steady mergers of small structures over time. Some galaxies achieve $\log M_{*}/M_{\odot}\sim11$ even in low density environments with active in situ star formation, but they come up against a steep barrier to growing further without mergers. Once galaxies are relegated to satellite status, in situ star formation becomes the dominant channel for increasing mass despite the gradual suppression of star formation activities due to environmental effects. 

Galaxy formation models are, of course, incomplete as yet and are unable to precisely describe the formation of stellar mass~\citep[e.g.][]{sparre15,furlong15,knebe15,song16}. Nonetheless, the success of the concordance $\Lambda$CDM cosmology gives credibility to the efforts to understand galaxy mass assembly, which is regarded as a consequential process of structure formation in the cosmological framework. This study demonstrates that similarly massive galaxies have a variety of evolution histories, depending on their halo environments. Meanwhile, empirical studies have found a considerable number of luminous galaxies in low density environments as well as dense environments~\citep[e.g.][]{croton05,khim15}. Motivated by these results, we will carry out a comparison study for massive galaxies that end up with similar properties but reside in different local halo environments.

\section*{acknowledgments}
We thank Rory Smith for his constructive comments and proofreading. We acknowledge the support from the National Research Foundation of Korea (NRF-2014R1A2A1A01003730). Numerical simulations were performed using the KISTI supercomputer under the programme of KSC-2014-G2-003. S.K.Y. acted as the corresponding author. This work was performed under the collaboration between Yonsei University Observatory and Korea Astronomy and Space Science Institute.


\end{document}